\documentstyle [12pt] {article}
\input epsf

\begin{document}
\oddsidemargin -.375in
\begin{flushright}
%ISU-NP-05-15\\
%KKU-NP-05-11\\
%December 2005\
\end{flushright}
\vspace {.5in}
\begin{center}
{\Large\bf Level statistics for two-dimensional oscillators\\}
\vspace{.5in}

{\bf A. Abd El-Hady ${}^{a,} \footnote{Permanent address : Physics
Department, Faculty of Science, Zagazig University, Zagazig, Egypt
}$ and A. Y. Abul-Magd ${}^b$\\} \vspace{.1in} ${}^{a)}$ {\it
Department of Physics, Faculty of Science, King Khalid University, Abha, Saudi Arabia}\\
\vspace{.1in} ${}^{b)}$ {\it
  Department of Mathematics, Faculty of Science, Zagazig University, Zagazig, Egypt}\\
\vskip .5in
%{\bf Abstract}

\bf{\today}

\end{center}

\begin{abstract}

We consider the level statistics of two-dimensional harmonic
oscillators with incommensurable frequencies, which are known to
have picket-fence type spectra. We propose a parametric
representation for the level-spacing distribution and level-number
variance, and study the variation of the parameters with the
frequency ratio and the size of the spectra. By introducing an
anharmonic perturbation, we observe a gradual transition to the
Poisson statistics. We describe the level spectra in transition
from harmonic to Poissonian statistics as a superposition of two
independent sequences, one for each of the two extreme statistics.
We show that this transition provides a suitable description for
the evolution of the spectrum of a disordered chain with
increasing long range correlations between the lattice sites.

PACS numbers: 03.65-w, 05.45.Mt, 05.30.

\end{abstract}

\newpage

%\twocolumn

%
%\begin{center}
\section{INTRODUCTION\label{intro}}
%\end{center}
%

Bohigas, Giannoni and Schmidt \cite{bohigas_1} have conjectured
that the spectral fluctuations of a quantum system whose classical
analog is chaotic follows the predictions of the random matrix
theory \cite{mehta_2}. This conjecture has been checked
numerically on a wide variety of systems (for recent reviews, see
e.g. \cite{guhr_3,mirlin_4}). The spectra of classically
integrable quantum systems are admitted as a sequence of
uncorrelated levels that follows the Poisson statistics. Berry and
Tabor \cite{berry_tabor_5} have given elaborate semiclassical
arguments in favor of this admission. The Poisson distribution for
regular spectra has been proven in some cases (see results by
Sinai \cite{sinai_6} and Marklof \cite{marklof_7}, for instance).

Not all of the integrable systems have a Poissonian level-spacing
distribution. The two-dimensional harmonic oscillator is a
classical example. Berry and Tabor \cite{berry_tabor_5} show that
the spacing distribution does not exist if the oscillator
frequencies are commensurable. It is peaked at a non-zero value if
the frequencies are incommensurable. The spectrum in this case has
nearly a picket-fence shape. The reason for this anomalous
behavior is the following. Berry and Tabor \cite{berry_tabor_5}
obtain the Poissonian level-spacing distribution by assuming
curved energy contours in the action space and applying the
stationary phase method. For the harmonic oscillator the energy
contours are flat and the method breaks down. Subsequent studies
of the energy spectra of harmonic oscillators are essentially a
continuation of \cite{berry_tabor_5}. Pandey and collaborators
\cite{pandey_bohigas_8,pandey_ram_9} use number theory to show
that harmonic oscillators have a strong level repulsion and no
fixed spacing distribution. A stable limit for the level-spacing
distribution is obtained using special averaging techniques by
Greenman \cite{greenman_10}. More recently, Chakrabati and Hu
\cite{chakra_11} demonstrate by numerical calculation that the
spacing distribution settles into a stable distribution as the
number of involved levels exceeds 5000. The existence and the
stability of the form of the spacing distribution of the
irrational two-dimensional oscillator call for further
investigations.

Because of its mathematical simplicity, the harmonic oscillator
provides solvable models in many branches of physics. It often
gives illustrations of abstract ideas. Nevertheless, the spectral
fluctuations of the oscillator have not been studied carefully
enough in the more than quarter-of-a-century that elapsed since
the publication of the work by Perry and Tabor. Recently,
picket-fence-like spectra have been observed in several elaborate
numerical experiments such as quantum graphs \cite{barra_12}. The
development of spectra of correlated disordered chains under the
influence of long range correlation, reported in
\cite{carpenna_13}, may be regarded as an evolution from the
Poissonian behavior towards that of a picket fence. This provides
an additional reason for revisiting the irrational two-dimensional
oscillator.

The aim of the present paper is to further explore the spectrum of
the two-dimensional oscillator, particularly when it is subject to
a small departure from harmonicity. Section \ref{2-D-oscillator}
is devoted to the numerical investigation of the spectral
fluctuations of the oscillators in terms of their level-spacing
distributions and level-number variances. Section \ref{transition}
shows that the violation of the harmonicity of the oscillators
results in a transition to the Poissonian statistics. Section
\ref{Analysis} describes the evolution of the spectra of a
one-dimensional lattice of a large number of sites interacting
according to a tight-binding hamiltonian under the influence of
long range correlations, observed in \cite{carpenna_13}, as a
transition from the Poissonian to the harmonic behavior. The
conclusions of this work are summarized in Section \ref{con}.

%\begin{center}
\section{The two-dimensional oscillator\label{2-D-oscillator}}
%\end{center}

In this section, we add a small contribution to the study of
energy spectra of two-dimensional harmonic oscillators, initiated
by Berry and Tabor \cite{berry_tabor_5}. The energy eigenvalues
for such a system are given by
\begin{eqnarray}
E_i^{harmonic} = \hbar \omega (n + \alpha m), \ \label{e_nm_1}
\end{eqnarray}
where $\omega$ is the larger frequency and $\alpha$ is the
frequency ratio while $i$ stands for the quantum-number pair $n$
and $m$.

We can represent each state by a point in the first quadrant of
the plane whose axes represent the quantum numbers $n$ and $m$.
For $\hbar \omega=1$, the number of states below some energy $E$
is equal to the number of points in the first quadrant below the
straight line cutting the $n$-axis at $E$ and the $m$-axis at
$E/\alpha$. Thus, the number of states below some energy $E$ is
proportional to $E^2$, the density of states increases linearly
with $E$, and the mean level spacing is proportional to $1/E$.

Although the two-dimensional harmonic oscillator is an integrable
system, its nearest neighbor spacing distribution does not follow
the Poisson distribution characteristic of generic regular
systems. Berry and Tabor \cite{berry_tabor_5} showed that the
spacing distribution $P(s)$ does not exist if $\alpha$ is a
rational number. For irrational values of $\alpha$, they
calculated 10 000 eigenvalues and constructed the histograms of
$P(s)$ for oscillators with $\alpha = 1/\sqrt{2}, 1/\sqrt{5},
e^{-1}$ and $\pi ^{-1}$. In the first three cases, the obtained
$P(s)$ is tightly peaked at $s = 1$ and shows weak dependence on
the frequency ratio. According to these authors, the spacing
distributions show similar behavior for $\alpha = 1/\sqrt{3},
1/\sqrt{7}$ and $(\sqrt{5}-1)/2$.  In the case of $\alpha=\pi
^{-1}$, however, the distribution is bimodal with peaks at $s$ =
0.5 and 4.3, which has been attributed to the close equality of
$\pi$ to many rational numbers.

In further studies of the irrational two-dimensional oscillator by
Pandey and collaborators \cite{pandey_bohigas_8,pandey_ram_9}, it
was shown, using number theory arguments, that in any integer
segment $[M,M+1]$ of the energy spectrum there are at most three
spikes in the spacing distributions of these integer segments.
They also argued that the spacing distributions do not reach
stability at all, while numerical studies by Chakrabati and Hu
\cite{chakra_11} suggest that stability is reached when the number
of states is about 5000.

\begin{figure}[h!tb]
\centerline{\epsfxsize=0.9\textwidth\epsfbox{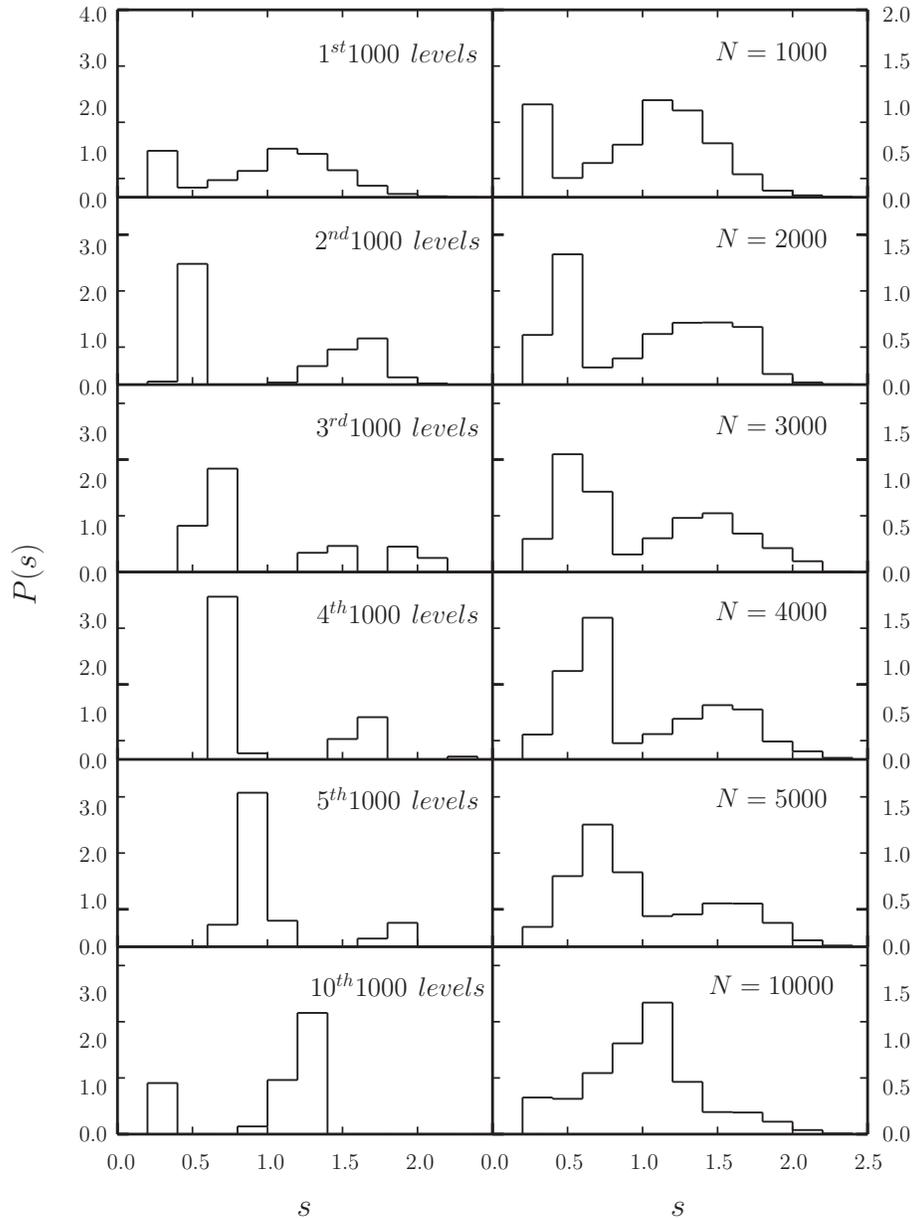}}
\caption{\label{fig1} \vspace{0.0cm}Left histograms : level
spacing distributions for 1000 successive levels. Right histograms
: level spacing distributions for the lowest 1000, 2000, ...,
10000 levels. All histograms are for the irrational oscillator of
Eq. (\ref{e_nm_1}) with $\alpha = 1/\sqrt{5}$. One should note
that the vertical left and right scales are different.}
\end{figure}

In Fig. \ref{fig1}, we show the spacing distributions for the
irrational two-dimensional oscillator with $\alpha = 1/\sqrt{5}$.
Left histograms show the spacing distributions for successive 1000
levels, while the right ones show the spacing distributions for
the lowest 1000, 2000, ..., and 10000 levels. The left histograms
of Fig. \ref{fig1} show the spike structure
\cite{pandey_bohigas_8,pandey_ram_9} even though 1000 states cover
more than one unit segment of the energy spectrum. The $N=5000$
and $N=10000$ histograms of the right part of Fig. \ref{fig1} show
that the spacing distribution does not reach stability when the
number of states reach 5000 as claimed by Chakrabati and Hu
\cite{chakra_11}.

Berry and Tabor \cite{berry_tabor_5} stated that they were not
able to prove that for any irrational oscillator that the level
spacing distributions settle into a stable form, but they also
stated that their numerical results suggest that they do.

Our numerical results displayed in Fig. \ref{fig1} show that the
positions of the spikes in the spacing distributions change from
one interval to another. Adding the contributions from different
intervals, we notice that the spike structure disappears and the
fluctuations in the spacing distributions may not qualitatively
change the shape of the distribution for sufficiently large $N$.
We also notice that the spacing distribution tends to be peaked
around $s=1$.
%fig1

\begin{figure}[h!tb]
\centerline{\epsfxsize=0.9\textwidth\epsfbox{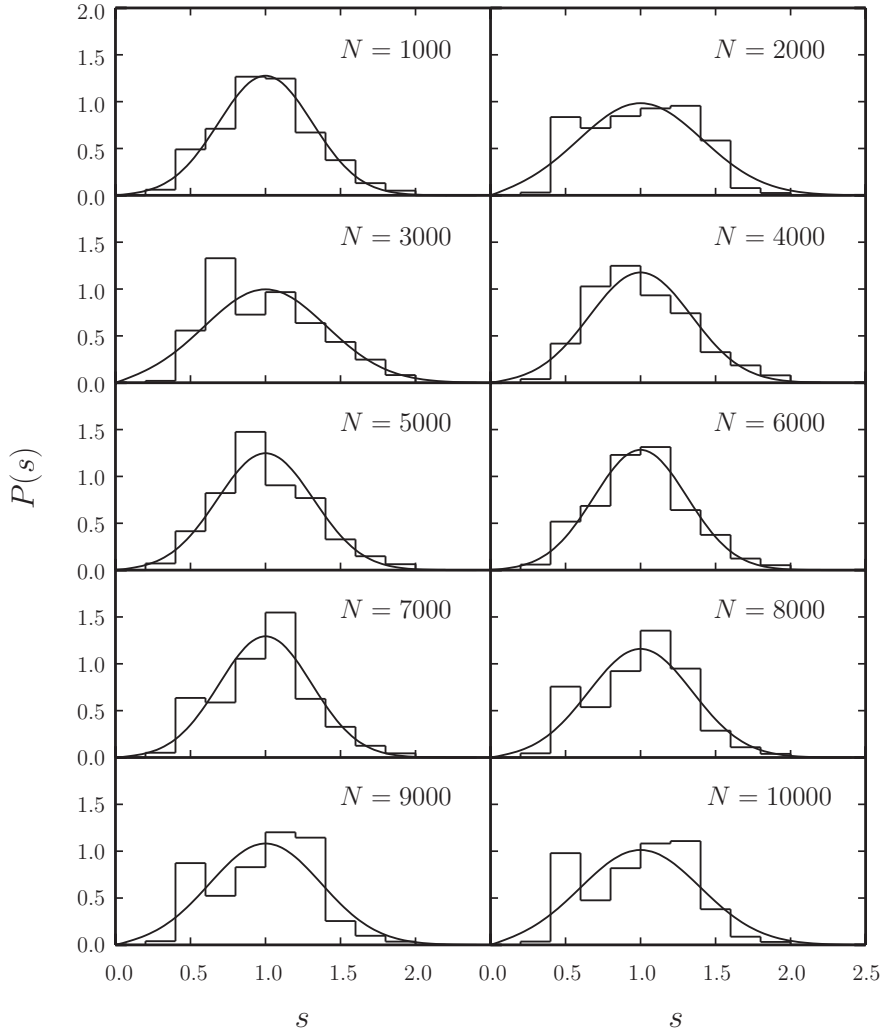}}
\caption{\label{fig2} \vspace{0.0cm}Level spacing distributions
for the first 1000, 2000, ..., 10000 levels of the irrational
oscillator of Eq. (\ref{e_nm_1}) with $\alpha = 1/\sqrt{2}$,
fitted to the modified Gaussian distribution of Eq.
(\ref{p(sw)_2}).}
\end{figure}

\begin{figure}[h!tb]
\centerline{\epsfxsize=0.8\textwidth\epsfbox{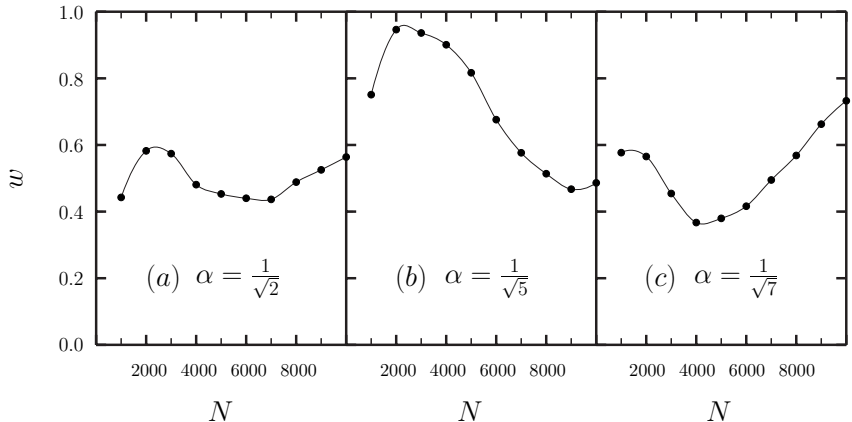}}
\caption{\label{fig3} \vspace{0.0cm}Dependence of the parameter
$w$ of the modified Gaussian distribution of Eq. (\ref{p(sw)_2})
on the size of the spectra for $\alpha = 1/\sqrt{2}$,
$1/\sqrt{5}$, and $ 1/\sqrt{7}$.}
\end{figure}

We would like to test the dependence of spectral fluctuations on
the number $N$ of levels constituting the spectrum and the
frequency ratio $\alpha$. We show the results for $P(s)$ for the
first $N$ levels of the oscillators with $\alpha = 1/\sqrt{2}$ in
Fig. \ref{fig2}, where $N$ is increased from 1000 to 10 000 by
steps of 1000. The spectra are unfolded using a cubic polynomial.
They are fitted to a modified Gaussian distribution

\begin{eqnarray}
P(s,w)=\frac{1}{\sqrt{\pi} w
{\rm{Erf}}(1/w)}\left[e^{-(s-s_c)^2/w^2}-e^{-(s+s_c)^2/w^2}\right],
\ \label{p(sw)_2}
\end{eqnarray}
where {\rm{Erf($x$)}} is the error function, $w > $ 0. This
distribution vanishes at the origin and has a mean value of $s_c$.
The pre-factor takes care of the normalization condition. In our
numerical analysis, we fixed the mean spacing at $s_c$ = 1. The
widths, however, are allowed to vary.
%fig2
We have also done the calculation for other values of $\alpha =
1/\sqrt{5}, 1/\sqrt{7}$. The best-fit values of the width
parameter $w$ are given in Fig. \ref{fig3}. The figure suggests
that the values of $w$ show regular oscillation with increasing
$N$ about a mean value that depends on the frequency ratio. The
mean values and the standard deviations are $w$ = 0.50 $\pm$ 0.06,
0.71 $\pm$ 0.18, 0.52 $\pm$ 0.12, for $\alpha = 1/\sqrt{2},
1/\sqrt{5}, 1/\sqrt{7}$, respectively. Nevertheless, we can say
that the best-fit values of $w$ agree within the error bars which
are less than~25$\%$.

We have also evaluated the number variance $\Sigma_2$ for the same
spectra. The level number variance is given by
\begin{eqnarray}
\Sigma_2(r) = <n^2(r,x)> - \left<n(r,x)\right>^2, \
\label{sigma_2_3}
\end{eqnarray}
%fig3
\\
where $n(r,x)$ counts the number of levels in the interval $[x, x
+ r]$ on the unfolded scale. The angular brackets in Eq.
(\ref{sigma_2_3}) denotes the averaging over the initial energies
$x$.

\begin{figure}[h!tb]
\centerline{\epsfxsize=0.85\textwidth\epsfbox{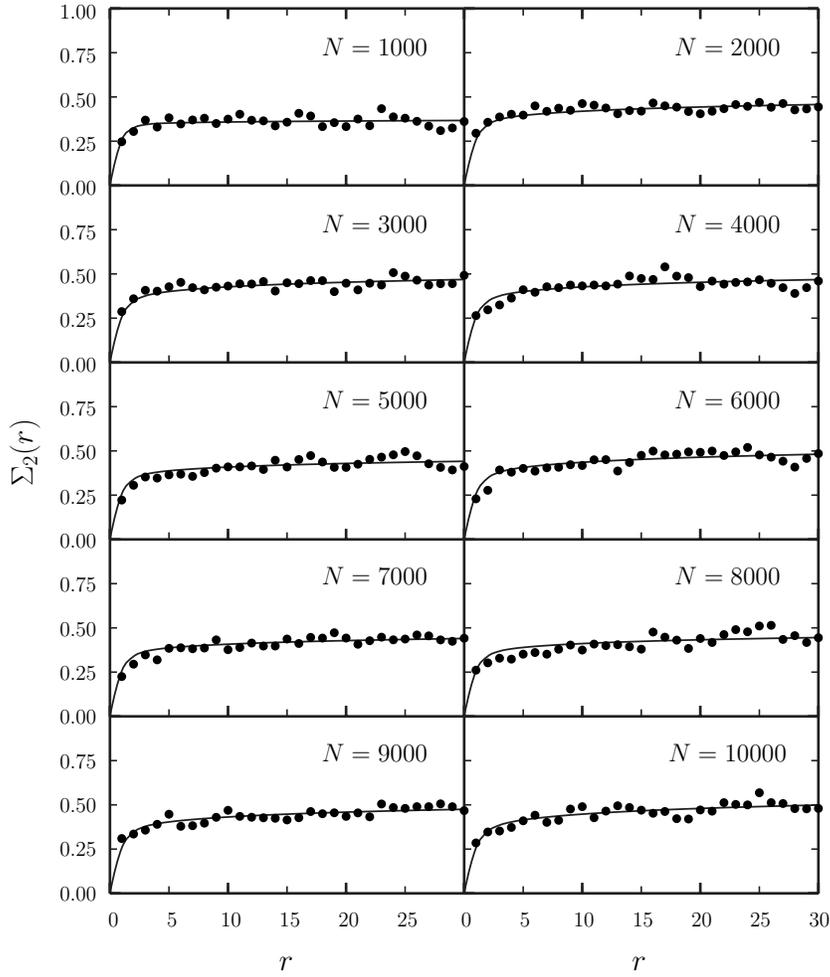}}
\caption{\label{fig4}\vspace{0.0cm}Level number variance
$\Sigma_2(r)$ for the first 1000, 2000, ..., 10000 levels of the
irrational oscillator of Eq. (\ref{e_nm_1}) with $\alpha =
1/\sqrt{2}$ fitted to Eq. (\ref{sigma_2_4}). }
\end{figure}

\begin{figure}[h!tb]
\centerline{\epsfxsize=0.85\textwidth\epsfbox{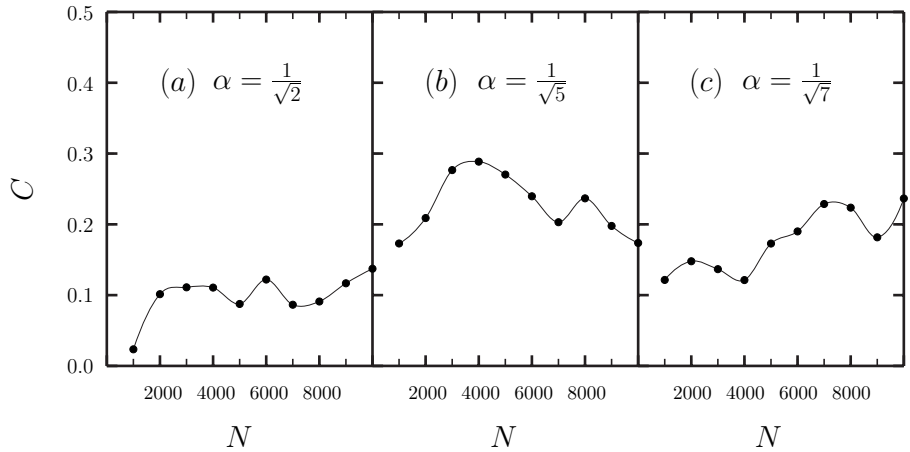}}
\caption{\label{fig5}\vspace{0.0cm}Dependence of the parameter $C$
of Eq. (\ref{sigma_2_4}) for the level number variance
$\Sigma_2(r)$ on the size of the spectra, in which $A$ = 0.34 and
$B$ =1.4 for $\alpha = 1/\sqrt{2}$, $1/\sqrt{5}$, and
$1/\sqrt{7}$. }
\end{figure}

In Ref. \cite{abdelhady_14} the spectral rigidity for a Gaussian
orthogonal ensemble of random matrices was successfully fitted by
a three-parameter formula, which was designed in accordance with
the asymptotic form of the exact expression. Inspired by this
success, we parameterize the level-number variance for the
irrational oscillator in the form

\begin{eqnarray}
\Sigma_2(r) = A ( 1 - e^{-B r} )( 1 + C \  {\rm{ln}}
 (r)). \ \label{sigma_2_4}
\end{eqnarray}
In Fig. \ref{fig4}, we use this formula to study the number
variance of levels of irrational oscillators with frequency ratio
$\alpha = 1/\sqrt{2}, 1/\sqrt{5}, 1/\sqrt{7}$.
%fig4
We have fixed the values of the two parameters $A$ = 0.34 and $B$
=1.4 and allowed the third parameter $C$ to vary with the size $N$
of the spectra. The variation of $C$ with $N$ is shown in Fig.
\ref{fig5} and is consistent with the variation observed in Fig.
\ref{fig3} for the width $w$ of the level-spacing distribution.

%fig5

%\begin{center}
\section{Transition to Poisson statistics\label{transition}}
%\end{center}

In this section, we consider the transition of the spectrum from
the harmonic behavior described above to the Poisson statistics.
Berry and Tabor \cite{berry_tabor_5} considered the level-spacing
distribution of a particle in a two-dimensional square-well
potential has a Poisson distribution. One can think of several
potentials that interpolate between the harmonic oscillator and
the square well. The mere existence of such interpolating
potentials suggests the possibility of a gradual transition from
the harmonic picket-fence shaped spectra to the spectra described
by the Poisson statistics.

%\begin{center}
\subsection{Model spectrum\label{model_spectrum}}
%\end{center}

As an example, one may consider the two-dimensional potential

\begin{eqnarray}
V(x,y) = V_0
\left[\left|\frac{x}{a}\right|^{\frac{2}{{(1-\gamma)}}} +
\left|\frac{y}{b}\right|^{\frac{2}{{(1-\gamma)}}}\right], \
\label{v_5}
\end{eqnarray}
which is a harmonic-oscillator potential if $\gamma$ = 0 and a
square well if $\gamma$ =1. This potential is a sum of two
components, one depending on the variable $x$, and the other
depending on $y$. The corresponding Schr\"{o}dinger equation in
cartesian coordinates allows the separation of variables. The
eigenvalues are, therefore, given by a sum of the eigenvalues of
the two components of the potential

\begin{eqnarray}
E(n,m) = E_1(a,n) + E_1(b,m). \ \label{e_6}
\end{eqnarray}
Our purpose is to perform a statistical analysis of the spectra
which involves mainly large quantum numbers. We can safely apply
the WKB approximation and write
\begin{eqnarray}
\int_{-t}^{t}dx\sqrt{\frac{2\mu}{\hbar
^2}\left[E_1(a,n)-V_0\left|\frac{x}{a}\right|^{\frac{2}{{(1-\gamma)}}}\right]}
= \left(n+\frac{1}{2}\right)\pi, \ \label{int_7}
\end{eqnarray}
where $\pm t$ are the turning points, with
\begin{eqnarray}
t=a\left(\frac{E_1(a,n)}{V_0}\right)^{\frac{1-\gamma}{2}}. \
\label{t_8}
\end{eqnarray}
Carrying out the integration, one obtains
\begin{eqnarray}
E_1(a,n) = V_0 \left[   \sqrt{\frac{\pi \hbar ^2}{2 \mu V_0
a^2}}\frac{   \Gamma\left(\frac{4-\gamma}{2}\right)   }{
\Gamma\left(\frac{3-\gamma}{2}\right) }(n+\frac{1}{2})
\right]^{\frac{2}{2-\gamma}}. \ \label{e_9}
\end{eqnarray}
It is easy to see that Eq. (\ref{e_9}) yields the correct result
in the case of a harmonic oscillator potential, for which $\gamma$
= 0 and $V_0 = \frac{1}{2} m \omega^2 a^2$, as well as in the case
of an infinite square well of width $2a$ where $\gamma$ = 1.
Substituting Eq. (\ref{e_9}) into Eq. (\ref{e_6}) yields

\begin{eqnarray}
E(n,m) = E_0\left[ \left(n+\frac{1}{2}\right)^{\frac{2}{2-\gamma}}
+ \alpha \left(m+\frac{1}{2}\right)^{\frac{2}{2-\gamma}} \right],
\ \label{e_10}
\end{eqnarray}
where
%\begin{eqnarray*}
$E_0=V_0 \left[   \sqrt{\frac{\pi \hbar ^2 }{2 \mu V_0 a^2}}\frac{
\Gamma\left(\frac{4-\gamma}{2}\right)   }{
\Gamma\left(\frac{3-\gamma}{2}\right) }
\right]^{\frac{2}{2-\gamma}}\ \label{e_10_2}$
%\end{eqnarray*}
and
%\begin{eqnarray*}
$\alpha = \left(\frac{a}{b}\right)^{\frac{2}{2-\gamma}}. \
\label{e_10_3}$
%\end{eqnarray*}

%\begin{center}
\subsection{A model for the transition from the harmonic statistics to the Poissonian
\label{A model for the transition}}
%\end{center}

The level spectrum for a system undergoing a crossover transition
between two statistics was successfully described in
\cite{abdelhady_14,abulmagd_15,abulmagd_16} as a superposition of
two independent sequences; each one follows one of the two
statistics. The model was first proposed by Berry and Robnik
\cite{berry_robnik_17} to study the transition from the Poisson
statistics to that of a Gaussian Orthogonal Ensemble. In this
model one has to calculate the gap function for each sub-spectrum

\begin{eqnarray}
E_i(s) = \int_{s}^{\infty}dy\int_{y}^{\infty}dxP_i(x), \
\label{es_11}
\end{eqnarray}
where $P_i(s)$ is level spacing distribution of the sequence $i$.
Here we consider a superposition of two sequences. The first is
Poissonian, for which

\begin{eqnarray}
E_1(s) = e^{-s}. \ \label{es_12}
\end{eqnarray}
The second is harmonic with a spacing distribution represented by
Eq. (\ref{p(sw)_2}), for which

\begin{eqnarray}
E_2(s,w)&=&\frac{1}{2\  {\rm{Erf}}(\frac{1}{w})} \biggl\{  2  +
  \left( 1 - s \right) {\rm{Erf}} (\frac{1 - s}{w}) -
  \left( 1 + s \right) {\rm{Erf}} (\frac{1 + s}{w})
\nonumber \\
&&+\frac{w}{\sqrt{\pi}} \left[ e^{-\left( \frac{{\left( 1 - s
\right) }^2}{w^2} \right) } -
       e^{-\left( \frac{{\left( 1 + s \right) }^2}{w^2} \right) }
       \right]
          \biggr\}. \label{e_13}
\end{eqnarray}
The transition from a harmonic sequence to a Poissonian is
described by a spacing distribution

\begin{eqnarray}
P(s,w,f) = \frac{d^2}{ds^2}\left[   E_1((1-f)s)E_2(fs,w) \right],
\ \label{p_14}
\end{eqnarray}
with $f$ varying from 1 to 0.  Substituting Eq. (\ref{es_12}) and
Eq. (\ref{e_13}) into Eq. (\ref{p_14}), we finally obtain

\begin{eqnarray}
P(s,w,f)\!\!\!\!\!&=&\!\!\!\!\!\frac{e^{-(1-f)s}}{2 \sqrt{\pi} w
{\rm{Erf}}(\frac{1}{w})} \biggl\{ (1-f)^2\nonumber \\
&&\!\!\!\!+ \frac{1}{2\sqrt{\pi}w}\biggl[
2e^{-\frac{(1-fs)^2}{w^2}}  [f^2-(1-f^2)w^2]    +e^{-\frac{(1+fs)^2}{w^2}}  [2f^2+(1-f^2)w^2]\biggr]\nonumber \\
&&\!\!\!\!+\frac{1}{2}(1\!\!-\!\!f)\biggl[(1\!+\!f\!-\!fs\!+\!f^2s){\rm{Erf}}(\frac{1\!\!-\!\!fs}{w})
\!+\!(-1\!+\!3f\!\!-\!\!fs\!+\!f^2s){\rm{Erf}}(\frac{1\!\!+\!\!fs}{w}
 )\biggr]\biggr\}.
 \label{p_15}
\end{eqnarray}

%\begin{center}
\subsection{Numerical Calculation\label{Numerical Calculation}}
%\end{center}

\begin{figure}[h!tb]
\centerline{\epsfxsize=0.8\textwidth\epsfbox{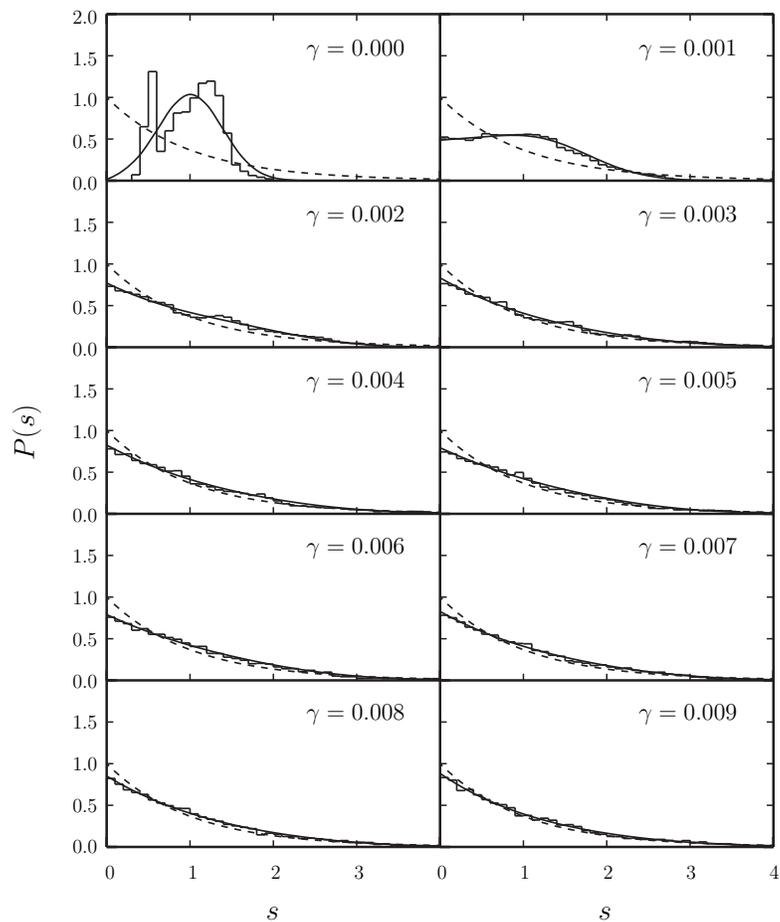}}
\caption{\label{fig6}\vspace{0.0cm}Level spacing distributions for
the first 10000 levels of the potential of Eq. (\ref{v_5}) with
$\alpha = 1/\sqrt{2}$, and different values of $\gamma$. The
irrational oscillator corresponds to $\gamma$ = 0. The solid and
dashed lines represent Eq. (\ref{p_15}) and the Poisson
distribution respectively.}
\end{figure}

\begin{figure}[h!tb]
\centerline{\epsfxsize=0.75\textwidth\epsfbox{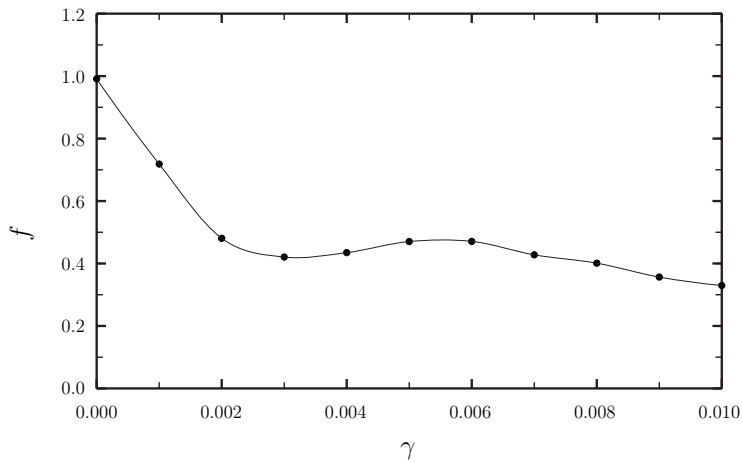}}
\caption{\label{fig7}\vspace{0.0cm}Dependence of the mixing
parameter $f$ of Eq. (\ref{p_15}), that fits the spacing
distribution of the potential of Eq. (\ref{v_5}) to the spacing
distribution for a superposition of a harmonic and a Poissonian
spectra, on the parameter $\gamma$ of the interpolating
potential.}
\end{figure}

Numerically, we use Eq. (\ref{e_10}) for $\alpha = 1/\sqrt{2}$ and
different values of the parameter $\gamma$ to calculate the first
10000 energy levels of the potential of Eq. (\ref{v_5}). We unfold
these spectra using a cubic polynomial and compare the spacing
distributions of the unfolded spectra with Eq. (\ref{p_15}). The
results are shown in Fig. \ref{fig6} and the best fit values of
the mixing parameter $f$ are shown in Fig. \ref{fig7}.

%fig6

%fig7

%\begin{center}
\section{Analysis of a numerical experiment on a physical system\label{Analysis}}
%\end{center}

The harmonic oscillator is a standard model for a large variety of
physical systems. It is natural to expect that the non-generic
behavior of the spectrum of the harmonic oscillator occurs in real
systems.

Carpena \emph{et al} \cite{carpenna_13} considered a
one-dimensional lattice of large number of sites $(N = 2^{17})$
interacting according to a tight-binding hamiltonian. They
introduced long range correlations between sites by allowing the
power spectral function to decay with exponent $\beta$. If $\beta$
= 0, the sites are uncorrelated and the spectrum has a Poisson
statistics. Increasing $\beta$, the level-spacing distribution
$P(s)$ acquired new functional forms essentially different from
the ones that occur in standard stochastic transitions form
Poisson to GOE statistics. They found a critical value of the
correlation exponent, $\beta_{cr}$ = 1.55 $\pm$ 0.05, beyond which
the Poissonian behavior is lost. As $\beta$ departs from
$\beta_{cr}$, the distribution $P(s)$ for small $s$ decreases and
simultaneously an increasing peak at $s$ = 1 grows. The spectrum
gradually develops a picket-fence behavior. This evolution was
also demonstrated by calculating the level-number variance
$\Sigma_2$ which acquired particularly small values for $\beta >
\beta_{cr}$.

\begin{figure}[h!tb]
\centerline{\epsfxsize=0.73\textwidth\epsfbox{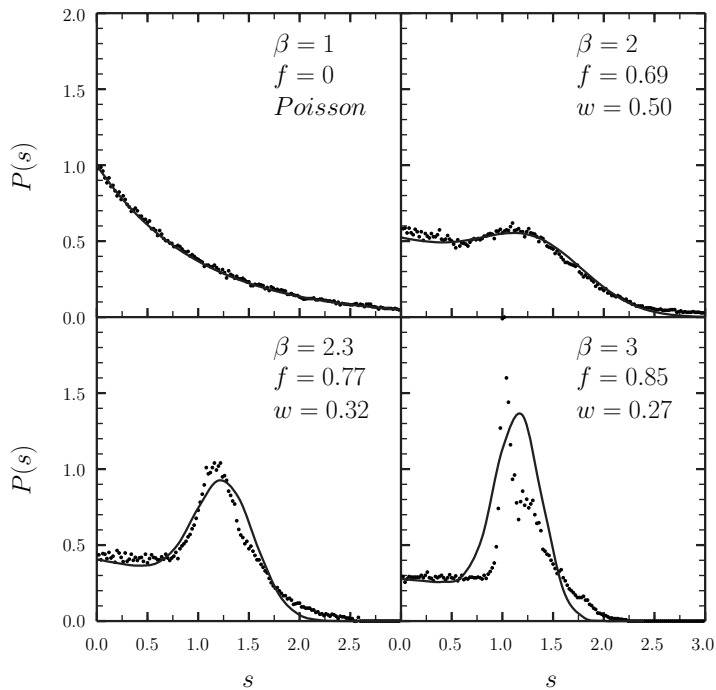}}
\caption{\label{fig8} \vspace{0.0cm}Level-spacing distribution for
correlated disordered chains considered in \cite{carpenna_13}
fitted to the distribution in Eq. (\ref{p_15}) for a superposition
of a Poissonian and an oscillator level sequences. }
\end{figure}

We show that the spectra of the disordered system obtained by
Carpena \emph{et al} \cite{carpenna_13} can be qualified as being
in transition from a Poisson statistics to that of a
harmonic-oscillator type. For this purpose, we fitted their
level-spacing distribution to that of a superposition of a
Poissonian and Gaussian functions, which is given by Eq.
(\ref{p_15}). In Fig. \ref{fig8}, we show the results for the
correlated disordered chains with the values of the correlation
exponent $\beta$, which have been considered in
\cite{carpenna_13}.
%fig8

\section{Conclusion\label{con}}
%\end{center}
More than a quarter of a century elapsed since Berry and Tabor
suggested that the spectra of irrational harmonic oscillators
suffer from a strong level repulsion. Most of the subsequent work
was based on number theory and showed that the spacing
distribution of the two-dimensional oscillator does not converge
to a stationary distribution. We have performed a numerical
analysis of the level-spacing distributions and level-number
variances for irrational oscillators with different frequency
ratios. We find that the parameters of these statistics oscillate
with increasing the size of the spectra around a mean value that
depends on the frequency ratio. We note, however, that the
amplitude of the oscillations and the difference between the mean
values are small enough to assume the presence of a new
``universality class" spectra which we may refer to as the
harmonic spectra. We then consider a two-dimensional potential
that interpolates between the oscillator and the infinite square
well; the latter has a spectrum satisfying the Poisson statistics,
and study the transition between the two classes of statistics. We
demonstrate that this transition exists in physical systems such
as disordered correlated chains.

%{\bf Acknowledgments}

\end{document}